# On the Cosmic Nuclear Cycle and the Similarity of Nuclei and Stars

**O. Manuel[1], Michael Mozina[2], and Hilton Ratcliffe[3]**

Repulsive interactions between neutrons in compact stellar cores cause luminosity and a steady outflow of hydrogen from stellar surfaces. Neutron repulsion in more massive compact objects made by gravitational collapse produces violent, energetic, cosmological events (quasars, gamma ray bursts, and active galactic centers) that had been attributed to black holes before neutron repulsion was recognized. Rather than evolving in one direction by fusion, nuclear matter on the cosmological scale cycles between fusion, gravitational collapse, and dissociation (including neutron-emission). This cycle involves neither the production of matter in an initial "Big Bang" nor the disappearance of matter into black holes. The similarity Bohr noted between atomic and planetary structures extends to a similarity between nuclear and stellar structures.

[1] Nuclear Chemistry, University of Missouri, Rolla, MO 65401 USA, om@umr.edu

[2]President, Emerging Technologies, P. O. Box 1539, Mt. Shasta, CA 96067, USA, *michael@etwebsite.com*
1-800-729-4445

[3] Astronomical Society of South Africa, PO Box 9, Observatory 7935, SOUTH AFRICA, ratcliff@iafrica.com

## I. INTRODUCTION

Hydrogen (H) and other lightweight elements are dominant on stellar surfaces and in the interstellar medium. Since the classical 1957 paper on element synthesis in stars by Burbidge *et al.* [1], it has been widely assumed that H-fusion is the main driving force for stellar luminosity and ordinary stellar evolution. The idea of a universe driven in one direction by H-fusion fits with the concept of H-production in an initial "Big Bang". However, a recent analysis of the systematic properties of all 2,850 known nuclides [2] revealed an even larger source of energy from repulsive interactions between neutrons in condensed nuclear matter [3-5].

Those results [3-5] and the abundances of isotopes and elements in meteorites, planets, the solar wind, the solar photosphere, and solar flares [6-9] showed that:

a) The Sun and other stars act as plasma diffusers, sorting lighter ions to their surfaces.
b) The interior of the Sun is made of common elements in rocky planets and meteorites – Fe, Ni, O, Si, and S – although the lightest elements (H and He) cover its surface.
c) Neutron-emission from the solar core, a neutron star, is the first step in a series of reactions that has steadily generated luminosity, neutrinos, solar mass fractionation, and an out-pouring of solar-wind hydrogen from the Sun over the past 4-5 Gy.
d) Neutron-emission from a neutron star is a statistical process, like the radioactive decay of ordinary nuclei *via* $\alpha$, $\beta$, $\gamma$, or spontaneous fission.

The Sun is an ordinary star, probably powered by the same processes as other stars. Prior to these recent papers [3-9], compact nuclear matter or black holes had been considered as the likely energy sources for the violent, more energetic events, like gamma ray bursts and quasars, but not as an energy source that might sustain luminosity in ordinary stars for billions of years.

However if the collapse of a neutron star is halted by neutron repulsion before becoming a singularity (a black hole), then repulsive interactions in super-massive neutron stars are the likely energy source that fragments cosmic matter to create clusters of galaxies, galaxies of stars, and planetary systems [10-14].

Before neutron repulsion was recognized as an energy source, Brown [10-14] noted evidence of repeated fragmentation in the cosmos, and Harutyunian [15] noted that the steady production of stellar luminosity and the violent fragmentation of matter into clusters of stars and galaxies are similar to the steady decay and the violent fragmentation of unstable nuclei [2, 15].

This effort to understand the nuclear cycle of the cosmos begins with our latest paper on the star next door. This paper [9] includes a few examples of the rigid, iron-rich structures that Mozina [16] noticed below the Sun's fluid photosphere in images from the SOHO and TRACE satellites. These satellite images of the Sun provide visual scientific evidence that falsifies the popular belief that the interior of the Sun consists mostly of H and He, like the solar atmosphere [17, 18]. Recent helioseismology data have now confirmed stratification at a relatively shallow depth beneath the visible photosphere, at about 0.5% solar radii (about 0.005 $R_o$) [19].

**INSERT FIGURE 1**

These new findings [3-9, 16, 19] and earlier ones [10-15, 17-18] lend credence to **a**.) the suggestion in the mid-1970s that the Sun formed on the collapsed core of the supernova (SN) that gave birth to the solar system [20-21], as shown in Fig. 1; **b**.) the finding that the Sun is iron-rich [6] like the terrestrial planets and contains proportionately less $^{136}$Xe and other r-products than the material that formed the giant Jovian planets from elements in the outer SN layers [22];

**c**.) the evidence against oscillations of solar neutrinos [23]; **d**.) Mozina's conclusion [16] of solar stratification and high levels of electrical and magnetic activity in the Sun's iron-rich stratified layers, **e**.) the suggestion that superfluidity of material in the interior of the Sun causes solar eruptions and climate changes [24], and **f**.) Birkeland's finding [25] at the start of the 20$^{th}$ century that many solar features resemble traits observed in the laboratory on magnetized metal spheres, including the link of the *aurora borealis* to solar magnetic activity.

**INSERT FIGURE 2**

Fig. 2 shows rigid structures in a "running difference" image of one small part of the Sun's iron-rich substructure, Active Region 9143, on 28 August 2000. The TRACE satellite generated this image using a 171 Å filter that is specifically sensitive to light emitted by Fe (IX) and Fe (X) iron ions. The team that operates the TRACE satellite system for NASA made a movie of the flare and mass ejection event that occurred from this Active Region 9143 on 28 August 2000. You can see the movie at http://trace.lmsal.com/POD/movies/T171_000828.avi or it is available here: http://vestige.lmsal.com/TRACE/Public/Gallery/Images/movies/T171_000828.avi.

For the present paper, chemical stratification and electromagnetic activity near the surface of the Sun [3-9, 16, 19, 24-25] are only of interest in demonstrating that some energy source other than H-fusion likely powers the Sun. We are more concerned here with the nuclear forces that have been able to sustain luminosity and an outpouring of hydrogen from the surface of the Sun and other stars over cosmological time scales of billions of years.

One external feature of the Sun, solar-induced variations in the geomagnetic field with a 2.65 h oscillation period, provided a hint almost three decades ago that the Sun itself might be a pulsar [26]. We will show below that repulsive interactions between neutrons in condensed nuclear matter is the driving force for a steady outpouring of hydrogen and luminosity from chemically stratified, iron-rich stars [3-9, 16, 19-21, 24-25] and a likely energy source for more chaotic events seen in the cosmos.

## II. THE ENERGY SOURCE FOR AN IRON-RICH, STRATIFIED SUN

A systematic enrichment of the lightweight isotopes of all five stable noble gases was recognized in the solar wind in 1983, extending over half of the entire mass range of the stable elements from A = 3 amu to 136 amu (atomic mass units) [6]. Other measurements [7-9] independently confirmed that the Sun selectively moves lighter ions into the photosphere, over the mass range of A = 25-207 amu [8], leaving little doubt that the interior of the Sun is iron-rich [6] like the material that formed iron meteorites and iron cores of rocky planets at the birth of the solar system [20-21]. Iron is however made of tightly packed nucleons [2] and is therefore an unlikely source of nuclear energy. This impasse lasted several years before it was realized in 2000 that repulsive interactions between neutrons in the solar core might be the source of both solar luminosity and the outpouring of solar-wind hydrogen from the surface of the Sun [3-5].

In the spring of 2000 five graduate students – Cynthia Bolon, Shelonda Finch, Daniel Ragland, Matthew Seelke, and Bing Zhang – who were enrolled in a graduate class entitled, *"Advanced Nuclear Chemistry: A Study of the Production and Decay of Nuclei"*, worked with the instructor, O. Manuel, to see if the properties of the 2,850 known nuclides [2] might reveal an unrecognized source of nuclear energy. Fig. 3 is a pictorial summary of the evidence they uncovered for repulsive interactions between neutrons in the nuclei of ordinary nuclear matter.

**INSERT FIGURE 3**

Data points in the drawing on the left side of Fig. 3 are experimental, but those on the right side at Z/A = 0 and Z/A = 1 are experimental only at A = 1, where they represent the neutron ($^1$n) and the lightest hydrogen isotope ($^1$H), respectively. The other data points on the right side of Fig. 3 were calculated from the mass parabolas defined by the mass data [2] at each value of A>1. Except for Coulomb repulsion, the n-n and p-p interactions are symmetric [3]. Values of the potential energy per nucleon (M/A) at Z/A = 0 and Z/A = 1 are therefore similar at low A.

As the value of A increases, values of M/A at Z/A = 1 become increasingly larger than those at Z/A = 0 because of the increasing contribution from Coulomb repulsion between positive charges. Thus, repulsive interactions between protons prevents the formation of massive cosmic objects compressed to nuclear density at Z/A = 1.

However at Z/A = 0 repulsive interactions between neutrons, documented by high values of M/A on the right side of Fig. 3, generate solar luminosity, energy in neutron stars, and an out-pouring of the neutron decay-product, $^1$H, from stars [3-5, 7-9]. Thus, in the stellar interiors

**(neutron-emission) + (neutron-decay) = (hydrogen production),**

despite reports that neutron stars are “dead” nuclear matter with each neutron in a neutron star having about 93 MeV less energy than a free neutron [27].

The idea of repulsive interactions between neutrons was revolutionary a few years ago [3], but Lunney *et al.* [28] have since agreed that useful information can be obtained by extrapolating nuclear mass data “*. . . out to homogeneous or infinite nuclear matter (INM)*” which “*. . . has a real existence, being found in the interior of neutron stars.*” [reference 28, p. 1042].

On the mass scale of ordinary nuclear matter, i.e., for A ┌ 300 amu, the interplay of repulsive and attractive interactions between nucleons [3] results in the following observations:

**a.)** Spontaneous alpha-decay in which heavy nuclei, near A ┌ 100 amu [2], emit packets of tightly packed nucleons as $^{4}He$;

**b.)** Spontaneous beta-decay, including electron-capture, at each value of A adjusts the charge density to the most stable value of Z/A [2];

**c.)** Spontaneous neutron-emission from neutron-rich nuclei over essentially the entire mass range of ordinary nuclear matter, e.g., $^{5}He$ – $^{149}La$ [2];

**d.)** A halo cloud of neutrons extending beyond the charge radius in light, neutron-rich nuclei, e.g., the two-neutron cloud at the surface of $^{11}Li$ outside the core nucleus of $^{9}Li$ [29];

**e.)** A rhythmic cycle in the potential energy per nucleon at Z/A = 0 (Fig. 3) over the entire mass range (A) caused by geometric changes in the packing of neutrons [4]; and

**f.)** Spontaneous fission of heavy nuclei with A ┌ 230 amu [2].

Coulomb repulsion becomes increasingly important in heavier nuclei, and neutrons are likely to be concentrated in the interior of these. This transition in the internal structure of ordinary nuclei seems to occur in the mass region where neutron-emission ceases and alpha-emission begins, near A ≈ 100-150 amu, [2]. For A >150 amu, the potential energy per nucleon (M/A) at Z/A = 0 (See right side of Fig. 3) gradually starts to increase with A [5], a trend used to extrapolate an upper limit of ≈ 22MeV on the excitation energy for a neutron in a neutron star [5].

The asymmetric fission of massive nuclei into heavy and light mass fragments with a mass ratio of ≈ 1.6/1 has long been linked to the closed shells of neutrons at N = 82 and 50, respectively. This empirical observation suggests that neutron repulsion in the core of heavy nuclei may contribute to nuclear fission. The occurrence of a similar fission process, like the above process **f.)**, in neutron stars might explain the universal occurrence of fragmentation in the production of solar systems, galaxies, and galactic clusters [10-15].

Properties **c.)–e.)** of ordinary nuclear matter may also be involved in the behavior of compact, neutron-rich stellar objects with A > ≈ $10^{57}$ amu ≈ 1 solar mass ($M_o$). Neutron-emission (process **c.**) is the process that sustains stellar luminosity over billions of years [3-9]. The presence of a halo cloud of neutrons around neutron stars (property **d.**) probably facilitates neutron-emission. Variations in values of the potential energy per neutron, M/A, with mass number, A, (property **e.**, See the right side of Fig. 3 at Z/A = 0) may cause super-massive neutron stars to mimic alpha-decay (process **a**) by emitting tightly packed clusters of neutrons as smaller neutron stars.

These processes would explain the analogous behavior noted between atomic nuclei and cosmic objects [15] and parallel the similarity Bohr [30] noted between atomic and planetary structures. However, another force comes into play on the cosmological mass scale that is unimportant on the mass scale of ordinary nuclear matter shown in Fig 3 – gravity.

## III. THE NUCLEAR CYCLE THAT POWERS THE COSMOS

It had long been assumed that gravitational collapse produces the neutron stars that are seen near the center of supernova (SN) debris. That is the scenario outlined in Fig 1 to explain the presence of a neutron star at the core of the Sun. There is indeed compelling evidence that highly radioactive debris of a supernova that exploded here 5 Gy ago formed the solar system before isotopes and elements from different SN layers had completely mixed [3-9, 20-21, 31-32].

However there are indications that the elements had also been sorted by mass [9] in the parent star (Fig. 1). This suggests that a carrier gas, an upward flow of $H^+$ ions generated by neutron-emission and neutron-decay, came from a neutron star at the core of the parent star prior to the SN explosion that produced the solar system (Fig. 1). Thus, the occurrence of a neutron star in the core of the Sun, in its precursor, and in other ordinary stars [3-9] implies that:

- **a.)** Stellar explosions may expose, but do not necessarily produce, the neutron stars that are seen in stellar debris; and
- **b.)** Neutron stars at the centers of ordinary stars were not made one-at-a-time in SN explosions but were more abundantly made in higher energy fragmentation events that produced our galaxy, probably in a high density region associated with active galactic nuclei (AGN), quasars, or massive neutron stars.

The origin of these high-density, energetic regions of space is not well known, e.g. [15, 33], but the link between high density and high cosmic activity suggests that gravitational collapse generates massive cosmic objects that are powered by repulsive interactions between neutrons. The remaining discussion will be based largely on the idea [33] that collisions and mergers of galaxies produce these high-density regions with high cosmic activity. Our conclusion about the cosmic nuclear cycle is independent of the process that generates these high density regions.

A recent review on galactic collisions notes that "transient galaxy dynamics", the recurrent collisions and mergers of galaxies, has replaced the classical view that galactic structures formed early in the universe and were followed by slow stellar evolution and the steady build-up of heavy elements [33]. Collisions or mergers of galaxies are highly prevalent, with ~1 in 10 of known galaxies engaged in some stage of physical interaction with another galaxy, and nearly all cohesively-formed galaxies, especially spirals, having experienced at least one collision in their lifetime. *"Galactic collisions involve a tremendous amount of energy. . . . . the collision energy is of order $10^{53}$ J. This is equivalent to about $10^{8-9}$ supernovae, . . ."* [reference 33, p. 6].

Harutyunian [15] notes that the compact nuclear objects produced by such high-energy events display many of the properties seen in ordinary nuclear matter, including the more rapid decay (shorter half-lives) of the more energetic nuclei [15].

Collisions are highly disruptive to all components of the galaxies, including the nucleus, and astronomers observe the collisional energy in many puzzling forms - quasars, gamma ray bursts, and active galactic centers (AGN). The extreme turbulence of active galactic nuclei (AGN) suggests the interactive presence of massive gravitational concentrations, possibly black holes [34] or super-massive neutron stars that fragment [10-15] into the multiple neutron stars that then serve as formation sites of new stars. Struck notes in the abstract of his review paper that *"Galactic collisions may trigger the formation of a large fraction of all the stars ever formed, and play a key role in fueling active galactic nuclei"* [reference 33, p. 1].

Matter is ejected from the massive object in the galaxy core in the form of jets, perhaps caused by an ultra-dense form of baryonic matter [35] in neutron stars or Bose-Einstein condensation of iron-rich, zero-spin material into a super-fluid, superconductor [24, 36] surrounding the galaxy core.

Hubble Space Telescope (HST) observations confirmed the hierarchical link suggested by Arp [37] between collisional systems and quasi-stellar objects - quasars. Quasars are frequently seen grouped, in pairs or more, across active galaxies, and are physically linked to the central galaxy by matter bridges. Isophote patterns indicate that the direction of motion of the quasars is *away* from their host galaxy, thereby stretching and weakening the matter bridge until the quasar separates completely. The implication is certain—quasars are physical ejecta from AGN, and become nuclei of nascent galaxies. The HST sightings *". . . provide direct evidence that some, and the implication that most, of the quasar hosts are collisional systems"* [reference 33, p. 105].

AGN, quasars, and neutron stars are highly prevalent, observable phenomena in all parts of the known universe. They have two significant properties in common: Exceptionally high specific gravity and the generation of copious amounts of "surplus" energy. In view of the repulsive forces recently identified between neutrons [3-5] and the frequency and products of galactic collisions [33], we conclude that neutron repulsion is the main energy source for the products of gravitational collapse.

Fig. 4 is a pictorial summary of the main features of the nuclear cycle that powers the cosmos: 1. Fusion; 2. Gravitational Collapse; and 3. Dissociation, including fragmentation and neutron emission.

**INSERT FIGURE 4**

## IV. CONCLUSION

Neutron-rich stellar objects produced by gravitational collapse exhibit many of the features that are observed in ordinary nuclei:

**a.)** Spontaneous neutron-emission from a central neutron star sustains luminosity and the outflow of hydrogen from the Sun and other ordinary stars;

**b.)** As a neutron star ages and loses mass, changes in the potential energy per neutron may cause instabilities due to geometric changes in the packing of neutrons (See the cyclic changes in values of M/A *vs.* A on the right side of Fig. 3 at Z/A = 0);

**c.)** Spontaneous fission may fragment super-heavy neutron stars into binaries or multiple neutron stars, analogous to the spontaneous fission of super-heavy elements; and

**d.)** Sequential fragmentation of massive neutron stars by emission of smaller neutron stars may resemble the sequential chain of alpha-emissions in the decay of U and Th nuclei into nuclei of Pb and He.

The nuclear cycle that powers the cosmos may not require the production of matter in an initial "Big Bang" or the disappearance of matter into black holes. The similarity Bohr noted in 1913 [30] between atomic and planetary structures extends to the similarity Harutyunian recently found [15] between nuclear and stellar structures. The recent finding [38] of a massive neutron star (CXO J164710.2-455216) in the Westerlund 1 star cluster where a black hole was expected observationally reinforces our doubts about the collapse of neutron stars into black holes.

Finally it should be noted that the elevated levels of $^{136}$Xe, an r-product of nucleosynthesis seen by the Gaileo probe into Jupiter [22], lend credence to Herndon's suggestion [39] that natural fission reactors [40] may be a source of heat in the giant outer planets.

**ACKNOWLEDGEMENTS**

Support from the University of Missouri-Rolla and permission from the Foundation for Chemical Research, Inc. (FCR) to reproduce figures from FCR reports are gratefully acknowledged. NASA and Lockheed Martin's TRACE satellite team made the image of rigid structures beneath the fluid photosphere and the movie of a solar flare coming from that region (Fig. 2). Cynthia Bolon, Shelonda Finch, Daniel Ragland, Matthew Seelke and Bing Zhang helped develop the "Cradle of the Nuclides" (Fig. 3) that exposed repulsive neutron interactions. Moral support is gratefully acknowledged from former UMR Chancellors, Dr. Raymond L.

Bisplinghoff (1975-1976) and Dr. Gary Thomas (2000-2005); former UMR Deans of Arts and Sciences, Dr. Marvin M. Barker (1980-1990) and Dr. Russell D. Buhite (1997-2002), and former UMR Chair of Chemistry, Professor Stig E. Friberg (1976-1979). This paper is dedicated to the memory of the late Professor Paul Kazuo Kuroda and his former student, Dr. Dwarka Das Sabu, for their deep personal commitment to the basic precepts of science and for their discoveries, including several papers [20-21, 31-32, 40] that contributed to the conclusions reached here.

## FIGURE CAPTIONS

**Fig. 1**. A scenario proposed in the mid-1970s to explain the unexpected link observed between specific isotopes of heavy elements with light element abundances in meteorites at the birth of the solar system [20-21]. According to this view, the Sun is iron-rich and formed on the collapsed core of a supernova (SN), material near the SN core formed iron cores of the planets near the Sun, and light elements from the outer SN layers formed the giant Jovian planets.

**Fig. 2.** A "running difference" image of the rigid, iron-rich structures beneath the photosphere in a small part of the Sun's surface revealed by the TRACE satellite using a 171 Å filter [16]. This filter is specifically sensitive to light emitted from Fe (IX) and Fe (X) iron ions. A movie made by the Lockheed Martin’s TRACE satellite team shows a solar flare and mass ejection (moving towards the upper left of the image) from this Active Region 9143 in 171Å light on 28 August 2000. The movie is here: http://trace.lmsal.com/POD/movies/T171_000828.avi, or it is available here: http://vestige.lmsal.com/TRACE/Public/Gallery/Images/movies/T171_000828.avi

**Fig. 3.** Left: The “Cradle of the Nuclides” is revealed when properties of ground-state nuclides [2] are plotted on a 3-D plot of mass per nucleon, M/A, *versus* charge density, Z/A, *versus* mass number, A. Right: For all values of A > 1 amu, mass parabolas defined by the data at other mass numbers intercept the front plane at (Z/A = 0, M/A = $(M/A)_{neutron}$ + ≈ 10 MeV) [3-5].

**Fig. 4.** The nuclear cycle that powers the cosmos: **1**. Fusion of lightweight nuclei, like $^{1}H$, into heavier ones, like $^{56}Fe$; **2**. Gravitational collapse of ordinary atomic matter into compact cosmological objects with Z/A = 0; and **3**. Neutron-emission (and neutron-decay) to produce the hydrogen fuel used in step 1. Charge density, Z/A, declines in steps **1.** and **2.** from Z/A =1 to Z/A = 0, but Z/A is unchanged in step **3.**, neutron-emission. Fragmentation [10-15] of super-massive neutron stars into smaller ones is not shown.

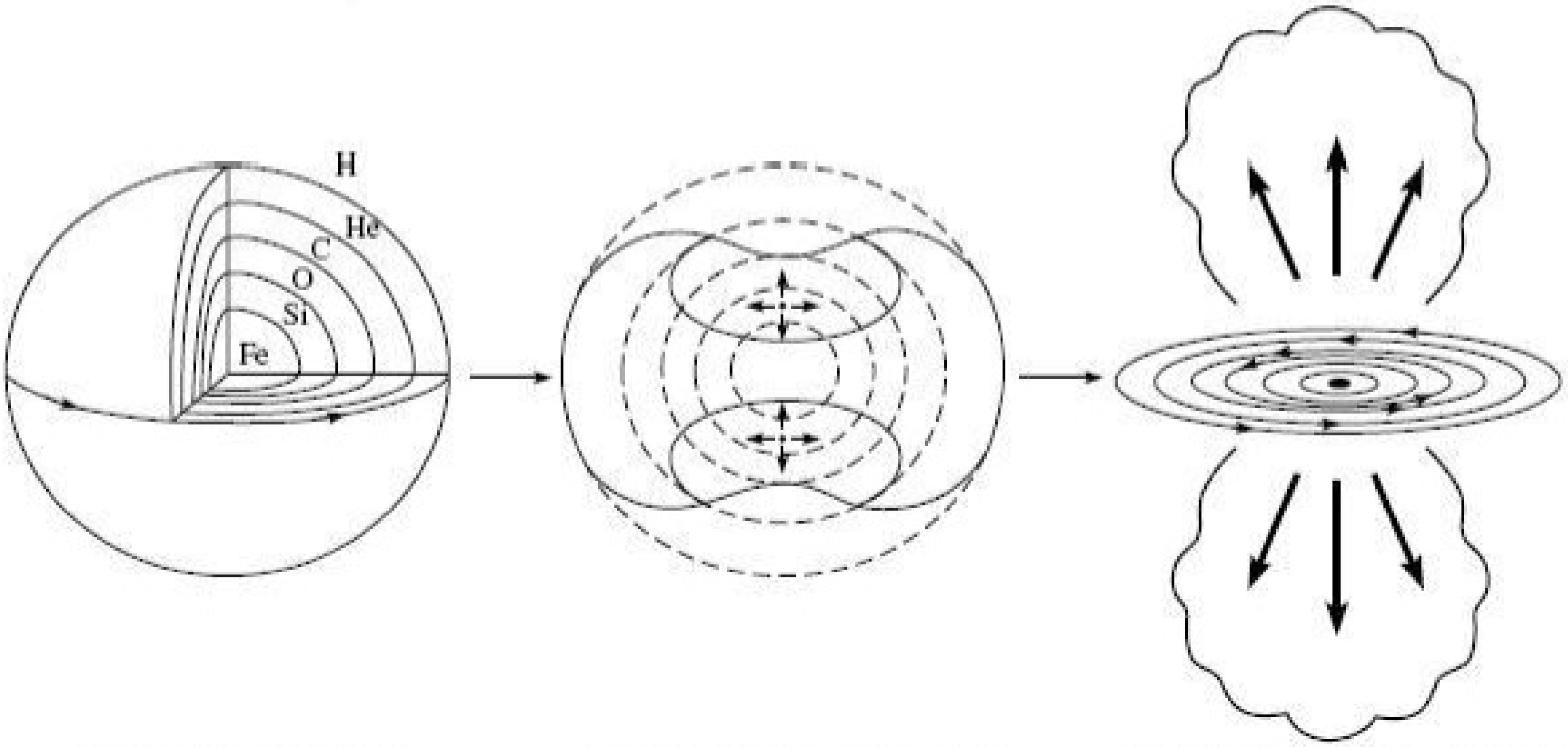

A massive spinning star becomes chemically layered near the end of its life, when asymmetric collapse occurs to conserve angular momentum.

The infall of low-Z elements causes an axially directed SN explosion, producing a rapidly expanding bipolar nebula with an equatorial accretion disk.

The Sun forms on the SN core; cores of inner planets form in the Fe-rich region around the SN core; Jovian planets form in the outer SN layers.

Figure 1

Manuel, Mozina, and Hilton

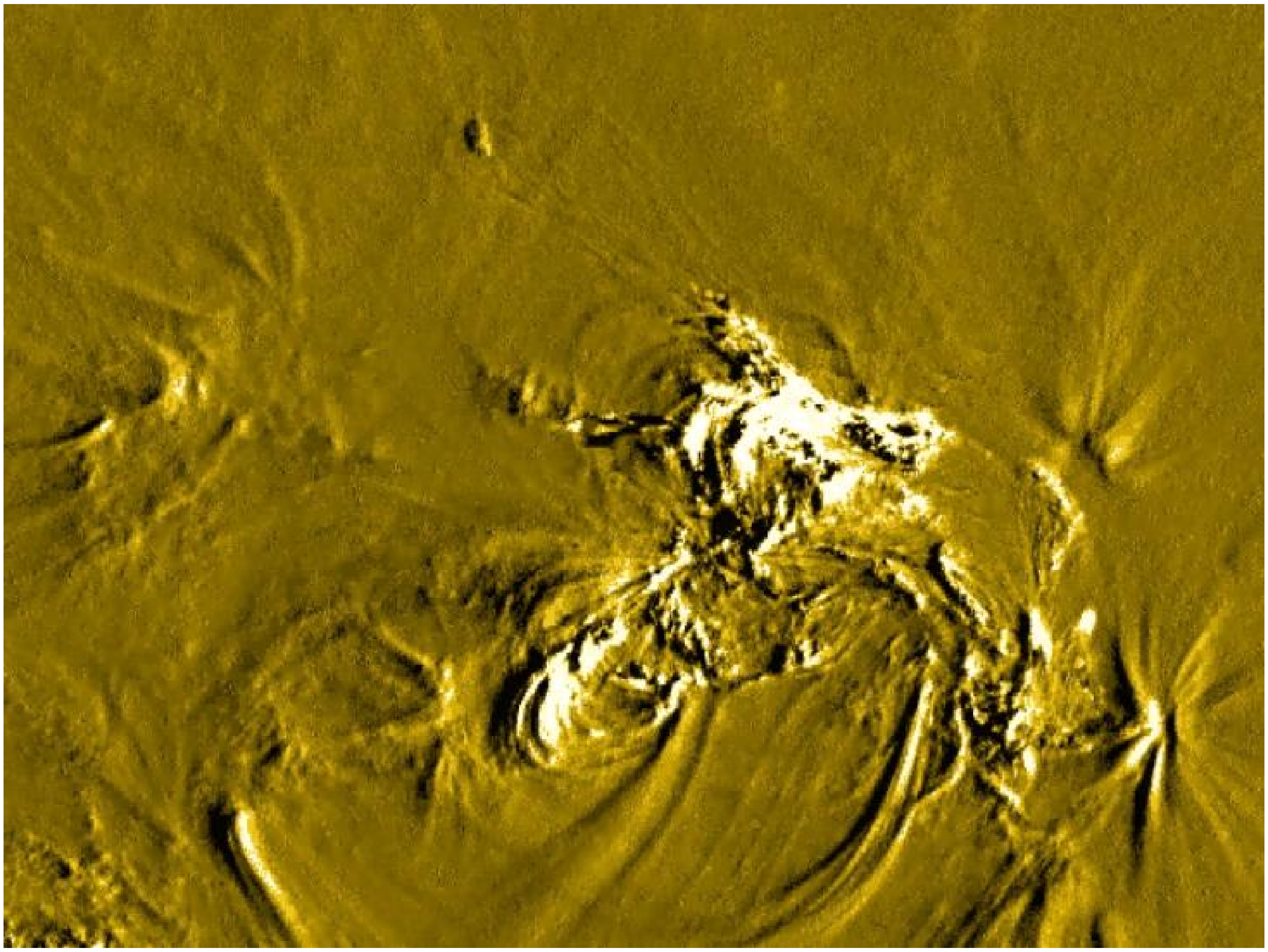

Figure 2

Manuel, Mozina, and Hilton

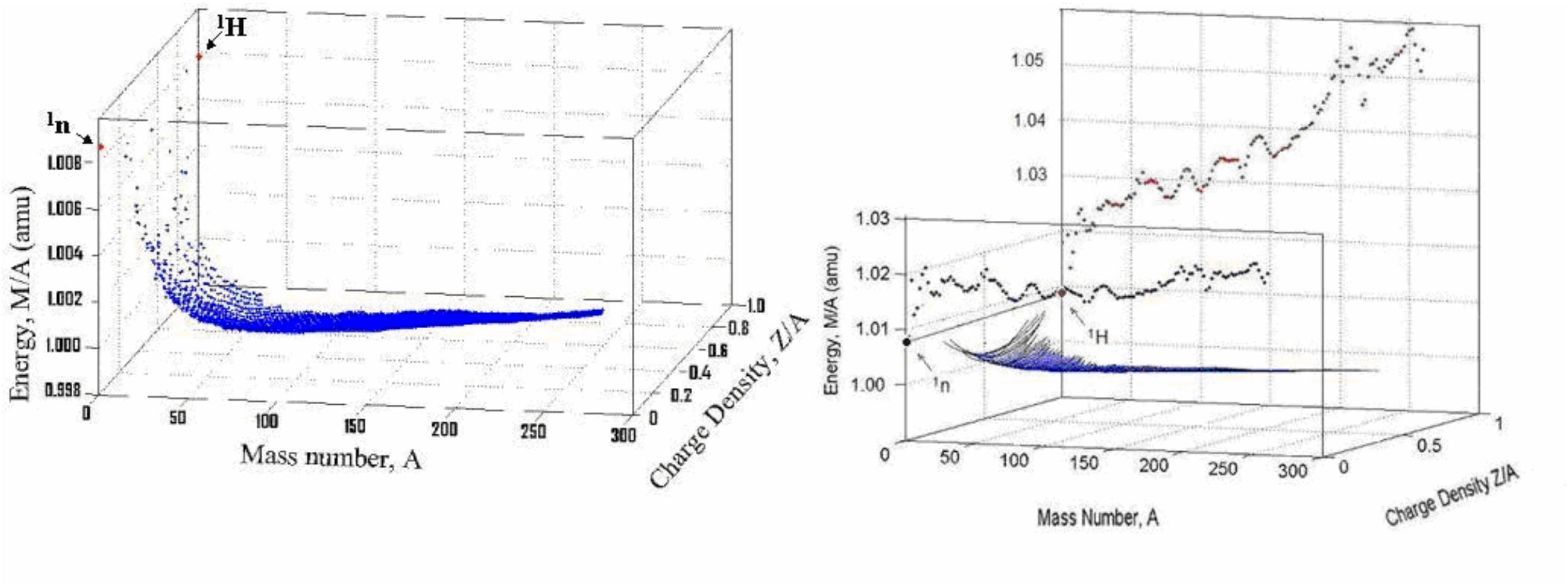

Figure 3

Manuel, Mozina, and Hilton

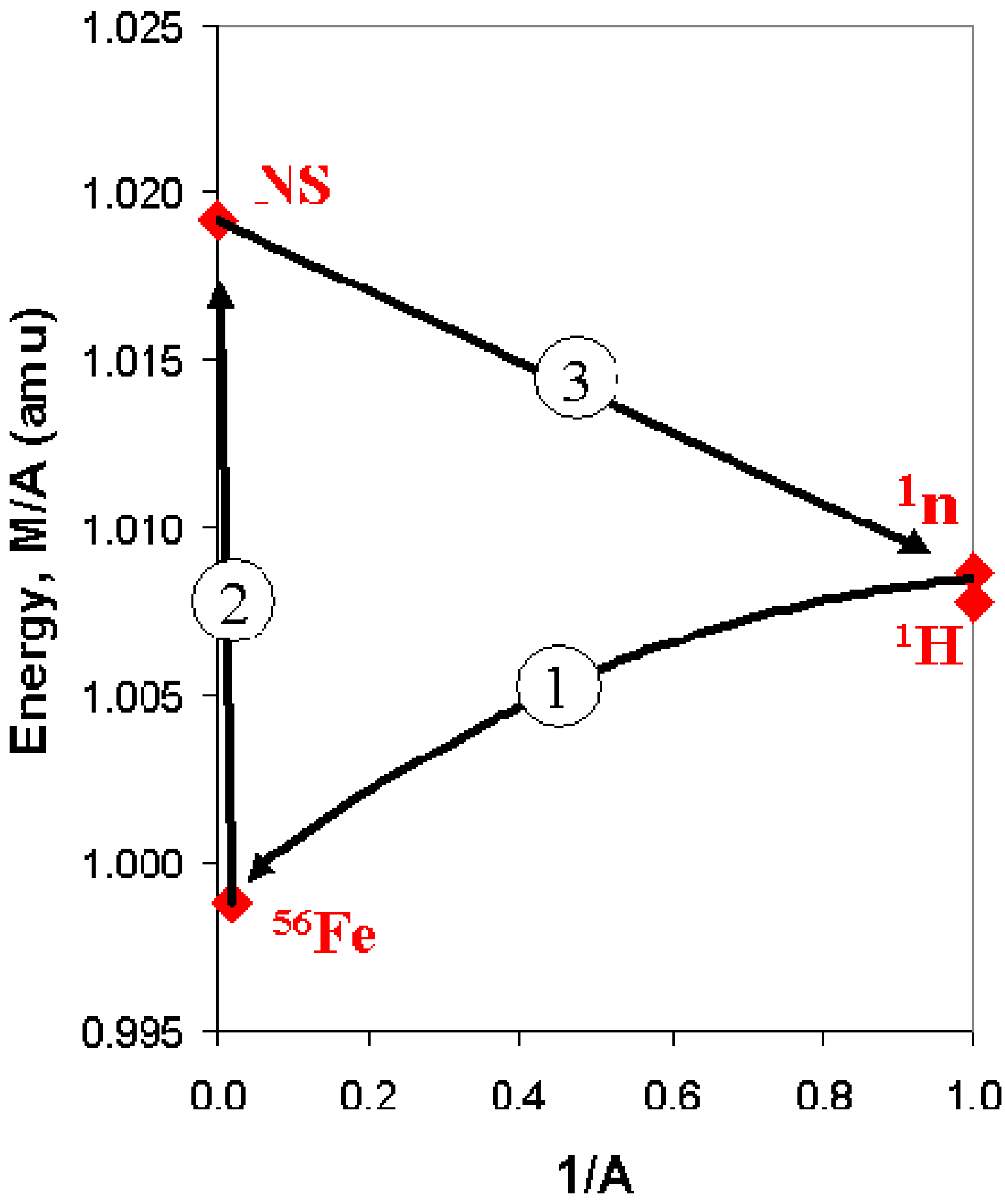

Figure 4

Manuel, Mozina, and Hilton